# The Zones Algorithm for
# Finding Points-Near-a-Point or Cross-Matching Spatial Datasets

Jim Gray, María A. Nieto-Santisteban, Alexander S. Szalay
Microsoft, Johns Hopkins University
MSR TR 2006 52
April 2006

**Abstract:** Zones index an N-dimensional Euclidian or metric space to efficiently *support points-near-a-point* queries either within a dataset or between two datasets. The approach uses relational algebra and the B-Tree mechanism found in almost all relational database systems. Hence, the Zones Algorithm gives a portable-relational implementation of points-near-point, spatial cross-match, and self-match queries. This article corrects some mistakes in an earlier article we wrote on the Zones Algorithm and describes some algorithmic improvements. The Appendix includes an implementation of *point-near-point*, *self-match*, and *cross-match* using the USGS city and stream gauge database.

## 1. Introduction

The article "There Goes the Neighborhood: Relational Algebra for Spatial Data Search" [1] introduced three spatial indexing methods for points and regions on a sphere: (1) *Hierarchical Triangular Mesh* (HTM) that is good for point-near-point and point-in-region queries with high dynamic range in region sizes, (2) *Zones* which is good for point-near-point queries with know radius, and also batch-oriented spatial join queries, and (3) *Regions* which is an algebraic approach to representing regions and doing Boolean operations on them, and answering point-in-region queries.

We have used all three methods extensively since that article was written [2], [4], [5], [6], [7]. The Zone Algorithm is particularly well suited to point-near-point queries with a search radius known in advance. However, when the radius is more than ten times larger than the zone height, the Zone Algorithm is less efficient. This high dynamic range is common in adaptive mesh simulations and many other spatial applications. In those cases, the HTM approach or perhaps DLS [3] is a better scheme. But, for many Astronomy and cartographic applications there is a natural scale (arcminute or mile) that covers many cases. For those applications, the Zone approach has some real advantages. First it works entirely within SQL, not requiring any extensions. So it is portable to any SQL database system. Second, it is efficient for batch-oriented spatial join queries – from 20 to 40 times faster than the object-at-a-time approach. The batch efficiency has given Zones a prominent place in building and using SkyServer.sdss.org and OpenSkyQuery.net. In particular, we use it to build the Neighbors table and to do batch-oriented cross match queries [4], [6], and [7].

This article corrects some subtle mistakes in the Zone algorithm described in [1], and presents some extensions. The basic changes are:
- The *radius-inflation* was wrong ($\theta' = \theta / cos(abs(dec))$ is wrong.) The correct computation affects both margin widths and search widths.
- The zone *margin* logic can be simplified.
- Self-match can do ½ the work by adding the symmetric pairs as a second step. Cross-match between different datasets doesn't have this symmetric option.
- A *Zone table* can eliminate the loop in multi-zone searches and matches.

**Terrestrial and Celestial Coordinates, A Rosetta Stone:**
Celestial coordinates are often expressed as Equatorial *right ascension* and *declination* (*ra, dec*) expressed in degrees. On the terrestrial sphere, people often use *latitude* and *longitude* (in degrees). This article uses (*ra, dec*) degrees, which geo-spatial readers will want to interpret as *lon/lat (not lat/lon)* – i.e. *lat ~ dec* and *lon ~ ra.* The code in the appendix uses (*lat, lon*) since the sample datasets are USGS places and stream gauges.

We also find it useful to represent (*ra,dec*) by their Cartesian coordinates – the unit vector $\boldsymbol{u} = (x,y,z)$ where $x$ points at the prime meridian equator, $z$ points to the north pole, and $y$ is normal to $x$ and $z$.

$$x = cos(dec)*cos(ra)$$
$$y = cos(dec)*sin(ra)$$
$$z = sin(dec)$$

## 2. Zone idea

All index mechanisms use a coarse index to produce candidate objects which are then filtered by some predicate (see Figure 1.) *Zones* uses a B-tree to bucket two-dimensional space (or higher-dimension space) to give dynamically computed bounding boxes (B-tree ranges) for spatial queries. A careful (expensive) geometry test examines all members of the bounding box and filters out false positives.

SkyServer.sdss.org uses the HTM package [2] to deliver a bounding box; but, calling the HTM has a drawback — SQL can evaluate a million spatial distance functions per second per cpu GHz while function calls to return sets of objects cost 100x more[1] than that. In particular, on a 1.8 GHz machine a table-valued function costs 200 µs per call and 26 µs per returned record. A typical call to the HTM routines or to the Zone-based `GetNearbyObjects()` routine described in Appendix (A.3) takes about 1400 µs to return 15 cities (there is actual computation in addition to the 590 µs fixed overhead suggested above). This 40:1 performance difference encourages using SQL operations as a coarse filter rather than using a user-defined function to give the bounding box. Pushing the logic entirely into SQL allows the query optimizer to do a very efficient job at filtering the objects. In particular, the Zone design gives a point-near-point performance comparable to the performance of the C# HTM sample code described in [5]. Both execute the following statement on the sample USGS Place table at a rate of about 600 lookups per second[2]:

```sql
select count(*) from  Place cross apply fHtmNearbyEq('P', lon, lat, 9).
```

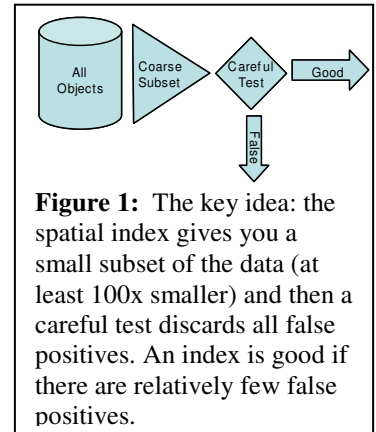

**Figure 1:** The key idea: the spatial index gives you a small subset of the data (at least 100x smaller) and then a careful test discards all false positives. An index is good if there are relatively few false positives.

The batch-oriented Zone algorithm gives a 34-fold speedup over calling a table-valued function for each neighbor computation. For the SkyServer load process, this turned a two-week computation into a 9 cpu-hour job that completes in less than an hour when run in parallel (see Section 4.3.)

The basic Zone idea is to map the sphere into *zones*; each zone is a declination stripe of the sphere with some *zoneHeight* (see Figure 2). For now, assume all zones have the same height. The zone just above the equator is zone number zero. An object with a declination of *dec* degrees is in zone:

$$zoneNumber = \lfloor dec / zoneHeight \rfloor \quad (1)$$

There are $\lceil 180/zoneHeight \rceil$ zones in all. The following code defines the `ZoneIndex` table.

```sql
create table ZoneIndex (
                zone int,                  -- the zone number
                objID bigint,              -- the object identifier
                ra float, dec float,       -- celestial coordinates
                x  float, y float, z float, -- Cartesian coordinates:
                                           --   for fast distance test
           primary key (zone, ra, objID))
```

The primary key index makes (`zone,ra`) lookups fast and clusters the zone bounding box elements.

The ZoneIndex table is populated from table T *approximately* as follows.
```sql
insert ZoneIndex
select floor( dec / @zoneHeight ), ra, dec, x, y, z                (2)
from T
```

---

[1] All performance measurements quoted here are from a Toshiba M200 computer with a 1.7 GHz Intel Celeron processor, 2MB L2 cache, 300MHz FSB, with 2GB PC2700 DRAM, 7200 rpm Seagate ST9100823A ATA disk, Windows XP SP2, and SQL Server 2005 SP1.

[2] The use of cross apply, which is part of the SQL standard, but only recently added to SQLserver is 30% faster than using a cursor to iterate over the objects.

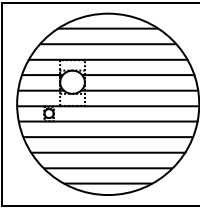

**Figure 2:** The division of the sphere into 12 zones (in practice there are thousands of zones). Two circular neighborhoods are shown, one inside a single zone (*minZone=maxZone*) and another crossing 3 zones (*minZone+2 = maxZone*.) The dotted boxes show how the *ra* filter and the *dec* filter further reduce the search. The *ra* search radius needs to be "expanded" by a *Alpha(θ,dec)* ~ *θ /cos(abs(dec))* (Section 2.1 defines *Alpha*).

If looking for all objects within a certain radius (*θ*) of point *(ra, dec)* then one need only look in certain zones, and only in certain parts of each zone. Indeed, to find all objects within radius *r* of point *ra, dec*[3], one need only consider zones between

         *maxZone* = ⌊ *(dec + θ)* / *zoneHeight*⌋                                            (3)
         *minZone* = ⌊ *(dec - θ)* / *zoneHeight*⌋

and within these zones one only need consider objects o with

         o.ra between *ra – Alpha(θ, dec)* and *ra + Alpha(θ, dec)*                (4)

There are some details that need extra mechanism. First, the *ra* search range within a zone must be expanded by *Alpha(ra,dec)* ~ *θ/cos(dec)*. Section 2.1 explains how to compute *Alpha*. Second, the *ra* range test of equation (4) must be computed modulo 360° to handle points near the prime meridian (*ra* = 0° or *ra* = 360°). Mechanisms to handle this are explained in section 2.2. As Section 2.3 explains, there are some subtle differences between self-match and cross-match. To simplify the discussion, Sections 2.1 through 2.3 only discuss one zone; Section 2.4 explains how multiple zone-searches are handled.

To give a preview, the *points-near-point* search, given *radius* = `@theta`, point = (`@ra`, `@dec`), `@alpha` = Alpha(`@theta,@dec`), where all angles are in degrees, first computes some preliminary values:

```
declare @ra float, @dec float, @theta float
select  @ra =237.5, @dec = 37.7,   -- San Francisco
        @theta = 4/60              -- 4 nautical miles radius (4 arcminutes)
-- Declare and compute the "working" variables x,y,z,alpha, zone
declare @x float, @y float, @z float, @zone int, @alpha float
Select  @x = cos(radians(@dec))*cos(radians(@ra)),
        @y = cos(radians(@dec))*sin(radians(@ra)),
        @z = sin(radians(@dec)),
        @alpha = dbo.Alpha(@dec,@theta),
        @zone = floor(@dec/@zoneHeight)
```

Then, using these parameters, the query to select objects nearby the point in the zone containing `@dec` is[4]:

```
select objID                                        -- return the objects
from   ZoneIndex                                    -- from ZoneIndex table
where  zone = @zone                                 -- in that zone number
  and ra  between @ra  – @alpha and @ra  + @alpha  -- quick filter on ra
  and dec between @dec – @theta and @dec + @theta  -- quick filter on dec
  and (x*@x + y*@y + z*@z) > cos(radians(@theta))  -- careful distance test
```

The following sections explain: (1) how to compute Alpha, (2) how to handle wrap-around at the meridian, and (3) how to look in all the relevant zones. But the basic logic is as simple as the single SQL statement above. This way of limiting the search is a typical bounding box approach but avoids calling an external procedure – it lets SQL do the math. The primary key on (`zone,ra`) makes this lookup very fast.

---

[3] We assume *ra* and *dec* have been normalized to ranges [0°, 360°] and [-90°, 90°] respectively.
[4] The mathematically correct *cos(θ) < **u·u**'* dot product is used here for clarity and shows the utility of the Cartesian coordinates (this is very efficient test). In practice, for small angles, cos(*θ*) is very close to 1.0 and the "significant" digits are 15 or more digits to the right (1-cos() is ~$10^{-15}$ for one arcsecond ~30 meters on Earth). To achieve high precision for small angles, the sin() calculation carries many more significant digits:
```
 and 4*power(sin(radians(@theta / 2)),2) >          -- careful distance test
        power(x-@x,2)+power(y-@y,2)+power(z-@z,2)   -- (2sin(r/2))^2
```

## 2.1 Alpha inflation near the poles

The zone algorithm described in [1] suggests that given a point *(ra, dec)* and a radius $\theta$,
(1) locate the zones implied by $dec - \theta$ to $dec + \theta$,
(2) compute the *inflated* radius $\alpha \sim \theta /\cos(abs(dec))$
     —this approximation is corrected in Section 2.2,
(3) then, for each zone look at the range  *ra - α* to *ra + α*.

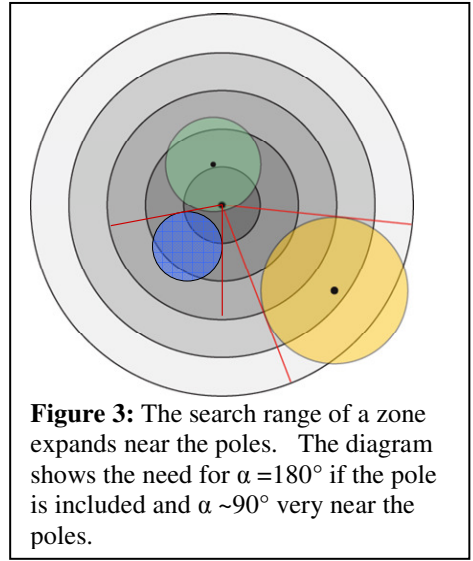

**Figure 3:** The search range of a zone expands near the poles. The diagram shows the need for $\alpha = 180°$ if the pole is included and $\alpha \sim 90°$ very near the poles.

This is approximately correct, but one needs to "inflate" α for search regions away from the equator. Using *cos(abs(dec))* as an inflator for zones between -80° and +80° is an acceptable approximation (relative error is less than $10^{-5}$). But when *abs(dec)*+ $\theta$ = 90°, α should be 90° and when *abs(dec)*+ $\theta$ > 90° (when the circle includes the pole as in Figure 3), α should be 180° so that the *ra* test for the pole zone will include points on the "far side" of the pole (see Figure 3).

Given a circle with opening angle $\theta$ around *(ra, dec)*, what are the limiting *ra* ranges of points on that circle? For simplicity, assume *ra = 0*. The *unit vector*, **u**, points at the circle center, the vector **n**, called the *northward vector*, is normal to **u** and together they define a plane including the north pole and the *westward vector* is **w**, as in the HTM paper [2].

$$\mathbf{n} = \begin{pmatrix} \cos dec \\ 0 \\ \sin dec \end{pmatrix}, \quad \mathbf{u} = \begin{pmatrix} -\sin dec \\ 0 \\ \cos dec \end{pmatrix}, \quad \mathbf{w} = \begin{pmatrix} 0 \\ -1 \\ 0 \end{pmatrix} \tag{5}$$

This defines a coordinate system centered on the sphere but with **u, w** defining the tangent plane at *(ra,dec)* and **n** is normal is to the tangent plane and points at *(ra,dec)*. Using the angle $\varphi$ running through the circle or radius θ around *(ra,dec)* to parameterize the points on the circle, the equation for points, **x**, on the circle is:

$$\mathbf{x} = \mathbf{n} \cos\theta + \mathbf{u} \sin\theta \cos\varphi + \mathbf{w} \sin\theta \sin\varphi \tag{6}$$

Substituting the definitions of the normal vectors (5) gives

$$\mathbf{x} = \begin{pmatrix} \cos\theta \cos dec - \sin dec \sin\theta \cos\varphi \\ -\sin\theta \sin\varphi \\ \cos\theta \sin dec + \sin\theta \cos dec \cos\varphi \end{pmatrix} \tag{7}$$

Using (7), for each point **x** on the $\theta$ circle, we can compute its right ascension α as:

$$\tan\alpha = \left(\frac{y}{x}\right) = -\frac{\sin\theta \sin\varphi}{\cos\theta \cos dec - \sin\theta \sin dec \cos\varphi} \tag{8}$$

Taking the derivative with respect to $\varphi$, in order to compute the extreme values gives:

$$-\frac{d\tan\alpha}{d\varphi} = \frac{\sin\theta \cos\varphi}{\cos\theta \cos dec - \sin\theta \sin dec \cos\varphi} - \frac{\sin\theta \sin\varphi \cdot \sin\theta \sin dec \sin\varphi}{(\cos\theta \cos dec - \sin\theta \sin dec \cos\varphi)^2} = 0 \tag{9}$$

Rearranging (9) and eliminating the denominator gives:
$\quad \sin\theta \cos\varphi(\cos\theta \cos dec - \sin\theta \sin dec \cos\varphi) - \sin^2\varphi \sin^2\theta \sin dec = 0$ (10)
Applying distribution gives:
$\quad \sin\theta \cos\theta \cos\varphi \cos dec - \sin^2\theta \sin dec \cos^2\varphi - \sin^2\theta \sin dec \sin^2\varphi = 0$ (11)
Dividing by sin $\theta$ and knowing $\sin^2\varphi + \cos^2\varphi = 1$, this simplifies to:
$\quad \cos\theta \cos\varphi \cos dec - \sin\theta \sin dec = 0$ (12)
Solving for $\cos\varphi$ and using tan = sin/cos:
$\quad \cos\varphi = \tan\theta \tan dec$ (13)

Equation (13) can be used to find the maximum value for tan $\alpha$ in equation (8). First refractor (8) to isolate the $\varphi$ terms by dividing the nominator and denominator of by cos $\theta$ cos *dec* and by using the *tan x= sin(x)/cos(x)* identity (three times):

$$\tan \alpha = \left| \left( \frac{\tan \theta}{\cos dec} \right) \left( \frac{\sin \varphi}{1 - \tan \theta \tan dec \cos \varphi} \right) \right| \tag{14}$$

Now substitute tan $\theta$ tan *dec* = cos $\varphi$ and use the $sin^2\varphi + cos^2\varphi = 1$ identity to get

$$\tan \alpha = \left| \left( \frac{\tan \theta}{\cos dec} \right) \left( \frac{\sqrt{1 - \cos^2 \varphi}}{1 - \cos^2 \varphi} \right) \right| = \left( \frac{\tan \theta}{\cos dec} \right) \left( \frac{1}{\sqrt{1 - \cos^2 \varphi}} \right) \tag{15}$$

Now substitute cos $\varphi$ = tan $\theta$ tan *dec*, from equation (12):

$$\tan \alpha = \left| \left( \frac{\tan \theta}{\cos dec} \right) \left( \frac{1}{\sqrt{1 - \tan^2 \theta \tan^2 dec}} \right) \right| \tag{16}$$

Using the *tan x= sin(x)/cos(x)* identity and rearranging terms, (16) can be rewritten as

$$\tan \alpha = \left| \frac{\sin \theta}{\sqrt{\cos^2 \theta \cos^2 dec - \sin^2 \theta \sin^2 dec}} \right| \tag{17}$$

And since cos(x+y) = cos(x)cos(y)-sin(x)sin(y) and cos(x-y) = cos(x)cos(y)+sin(x)sin(y) (17) simplifies to

$$\tan \alpha = \left| \frac{\sin \theta}{\sqrt{\cos(dec - \theta) \cos(dec + \theta)}} \right| \tag{18}$$

Solving for $\alpha$

$$\alpha = \left| \text{atan} \left( \frac{\sin \theta}{\sqrt{\cos(dec - \theta) \cos(dec + \theta)}} \right) \right| \tag{19}$$

There is a special case: when *abs(dec)*+ $\theta \geq 90°$, then $\alpha = 180°$.

To summarize, for $\theta < 1°$ and *abs(dec)* < 80°, the approximation $\alpha \sim \theta$ /cos(abs(dec)) has a relative error below $10^{-5}$. So, it is an adequate approximation for many terrestrial applications. But equation (19) should be used in general. In SQL the $\alpha$ computation is expressed as:

```
create function Alpha(@theta float, @dec float) returns float as
   begin
      if abs(@dec)+@theta > 89.9 return 180
      return(degrees(abs(atan(sin(radians(@theta)) /
                           sqrt(abs(   cos(radians(@lat-@theta))
                                     * cos(radians(@lat+@theta))
         )         )  )   )    )  )
   end
```

As a final note, the *Alpha(theta, dec)* computation computes *Alpha* for the entire circle which may touch many zones. The bounding box for these more distant zones are slightly too large. A more accuate $\alpha$ could be computed for each zone, in which case it would be smaller in zones away from the central *dec* zone – but that is a minor optimization. The calculation here is only slightly conservative.

## 2.2. Handling margins if there is wrap-around

Given a zone table for all cities, if we ask for places within 10 arc minutes of Greenwich, UK (*lat, lon*) = (51.48, 0) using a query like:
```sql
select objID                                       -- return the objects
from   ZoneIndex                                   -- from ZoneIndex table
where  zone = @zone                                -- in that zone number
  and ra between ra – @alpha and @ra + @alpha      -- quick filter on ra
  and dec between @dec – @theta and @dec + @theta  -- quick filter on dec
  and (x*@x + y*@y + z*@z) > cos(radians(@theta))  -- careful distance test
```
The query would not find London (about 5 arc minutes west of Greenwich.) Indeed the query does not find any place West of Greenwich since such places have *ra* (*longitude*) close to 360 ° rather than close to 0 ° (see Figure 4). This *spherical wraparound problem* requires that the *ra* test be done modulo 360°.

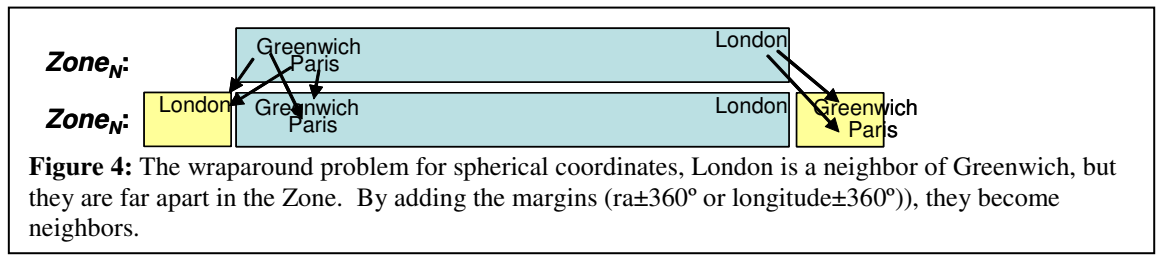

**Figure 4:** The wraparound problem for spherical coordinates, London is a neighbor of Greenwich, but they are far apart in the Zone. By adding the margins (ra±360º or longitude±360º)), they become neighbors.

The simplest solution to wraparound is to modify the query to be:
```sql
select objID                                       -- return the objects
from   ZoneIndex                                   -- from ZoneIndex table
where  zone = @zone                                -- in that zone number
  and (  ra between ra – @alpha and @ra + @alpha   -- quick filter on ra
      or ra between ra + 360 – @alpha and @ra + 360 + @alpha
      or ra between ra – 360 – @alpha and @ra – 360 + @alpha)
  and dec between @dec – @theta and @dec + @theta  -- quick filter on dec
  and x*@x + y*@y z*@z > cos(radians(@r))          -- careful distance test
```
This simple approach works, but it triples the index probes and so makes the query more expensive. Indeed, with SQLserver this query scans the whole zone rather than doing 3 probes because the optimizer is not smart enough to see the pattern. To "trick" the SQL Server optimizer into picking the correct plan, the above query must be expressed as the union of the three "between" predicates. Since at most two of the three probes are necessary, one can just have three SQL statements and guard two of them with if-statements so that a clause is invoked only if needed.

An alternate approach that trades disk storage space for simplicity and slightly better run-time performance replicates the *margin* objects in the `ZoneIndex` table as follows.
```sql
alter table ZoneIndex add margin bit not null default (0)
```
Then we add in the left and right margin objects (notice the margin Boolean is one).
```sql
insert ZoneIndex
  select zone, objID, ra–360, dec, x, y, z, 1
  from   ZoneIndex
  where ra >= 180                                  -- left margin
union
  select zone, objID, ra+360, dec, x, y, z, 1
  from   ZoneIndex
  where ra < 180                                   -- right margin
```

This doubles the size of the table; but, we assume that most of those marginal records will never be read from disk. If one knows that $\theta$ will be limited, then one need only replicate the *Alpha(@dec,@theta)* left and right margins. This is what we do for the SkyServer, assuming a ½ minute radius, so the margins increase the zone table by 0.001%.

With margins added to the `ZoneIndex` table, the original query works correctly near the prime meridian.

```sql
select objID                                          -- return the objects
from   ZoneIndex                                      -- from ZoneIndex table
where  zone = @zone                                   -- in that zone number
  and ra between ra - @alpha and @ra + @alpha         -- quick filter on ra
  and dec between @dec - @theta and @dec + @theta     -- quick filter on dec
  and (x*@x + y*@y + z*@z) > cos(radians(@theta))     -- careful distance test
```

## *2.3. Cross-Match and Self-Match*

Applications often want to find, for all objects, all *neighbors* within a certain radius – called a *self- match* of one dataset is compared with itself or a *cross-match* if correlating two different datasets. Zones are a good way of doing cross-match and self-match. In astronomy the neighborhood radius is often on the scale of 1 arcminute or less, while in terrestrial applications the radius is often 10x that (~10 nautical miles or more.)

The simplest way to compute this is to use the per-point logic of section 2.2 to define a function `GetNearbyObjects(@lat, @lon, @theta)`. Then, the self match is just

```sql
select ZI.objID as objID1, N.objID as objID2, distance
from ZoneIndex ZI cross apply GetNearbyObjects (lat, lon, @theta) N
where zi.margin = 0
  and ZI.objID!= N.objID
```

Similar logic works for cross-match. But, as explained earlier, the batch-oriented approach is twenty to forty times faster because it bypasses the individual function calls for each object. This section explains those optimizations.

When doing cross match in a zone, `@alpha` for all comparisons can be set to *maxAlpha*= Alpha($\theta$,*MaxDec*) wide where *MaxDec* is the max absolute value of *dec* for that zone. This is conservative, the bounding box will be a little too big in most cases, but it is a minor penalty to pay for the simpler design and it saves many *Alpha* computations.

Also, when doing cross-match or self-match, only the second dataset needs to have a margin – that is the first member of the match must be *native* but the second can be *marginal* (Figures 5 and 6.)

Next observe that self-match is symmetric. When matching a dataset with itself, if (obj1, obj2) are in the answer set, then (obj2, obj1) will be in the answer set as well (see Figure 5).

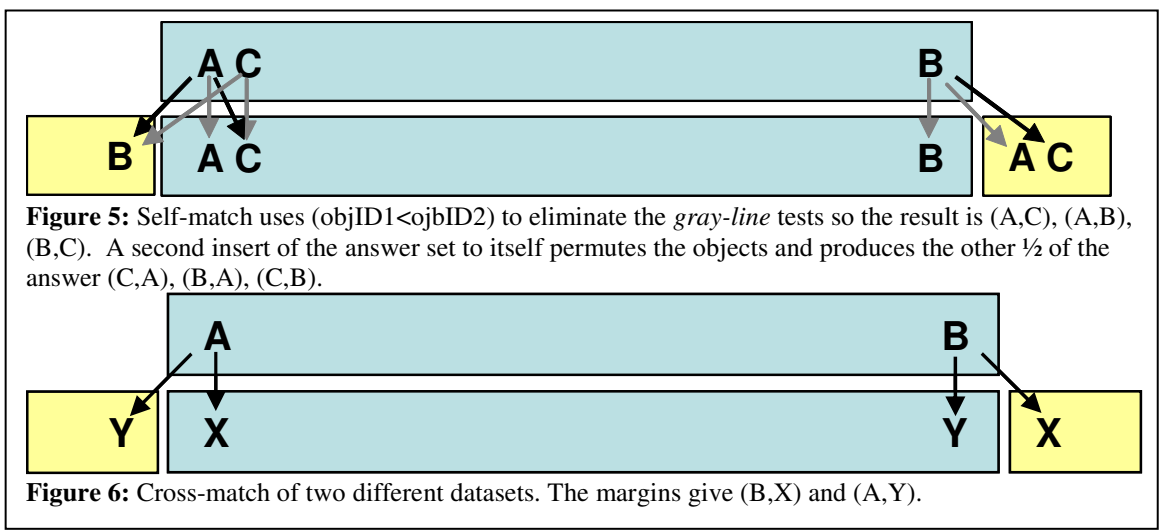

**Figure 5:** Self-match uses (objID1<ojbID2) to eliminate the *gray-line* tests so the result is (A,C), (A,B), (B,C). A second insert of the answer set to itself permutes the objects and produces the other ½ of the answer (C,A), (B,A), (C,B).

**Figure 6:** Cross-match of two different datasets. The margins give (B,X) and (A,Y).

So, cross-match of a zone with itself is:
```sql
select Z1.objID as ObjID1, Z2.objID as ObjID2  -- get object pairs
into #answer                                   -- put answer in a temp table
from   ZoneIndex Z1 join ZoneIndex Z2          -- from two copies of ZoneIndex
where  Z1.zone = Z2.zone                       -- where the two zones match
  and Z1.objID < Z2.objID                      -- only do ½ the expensive tests
  and Z1.margin = 0                            -- where first object is native
  and ra between ra -@maxAlpha and @ra + @maxAlpha -- quick filter on ra
  and dec between @dec - @theta and @dec + @theta  -- quick filter on dec
  and (cx*@x + cy*@y + cz*@z) > cos(radians(@theta))-- careful test
insert #answer                                 -- add the other ½ of the answers
select ObjID2, ObjID1                          -- permuting the object order
from #Answer                                   --  fetchign from answer table
```

There is also a self-match symmetry when matching two different Zones Z1 and Z2. One can match Z1 to Z2 and later match Z2 to Z1, or one can just match Z1 to Z2 and then add in the reversal { (z2,z1) | (z1,z2) ε Z1**X**Z2 } to the answer. In this way one need only exmine zones where Z1.zone<Z2.zone. These two symetries save approximately ½ the work in a self-match and speed the self-match computation by 30%.

These symmetries do not apply when cross-matching two different datasets. But, a technique that works for both self-match and cross-match is to include only native objects of the first dataset (see Figures 4, 5, 6.) Indeed, that restriction prevents duplicate entries in the answer set (e.g., preventing (B,A) from appearing twice in Figure 5 and (A,X) appearing twice in Figure 6.)

Since the radius, $\theta$, is known, the margin for a zone need only be *maxAlpha*. As mentioned above, this translates into a tiny increase in the number of rows in the zone table (0.001% for the SkyServer.).

Summarizing the optimizations for both self-match and for cross-match:
1. Each zone-zone comparison is a nested loops join that is cache and disk efficient.. Essentially it is a parallel sweep along the zone longitude of each zone comparing local values. It reads the disks sequentially and the data largely fits in the cpu cache.
2. Each zone-zone comparison is independent, so each can be done in parallel with the others or groups can be batched together. This parallelism can give huge speedups.
3. If {(zone1, zone2)} is the set of all comparisons, then for self-match one need only compare zones where zone1 ≤ zone2, saving ½ the search work. Cross-match needs to consider all pairs.

## 2.4. Searching multiple zones

For simplicity, the discussion so far has been in terms of a single zone. As Figures 2 and 3 show, a neighborhood may involve several zones. Indeed, for angle $\theta$ and declination *dec*, the search involves
```sql
zone between floor((@dec - @theta)/@zoneHeight)
         and floor((@dec + @theta)/@zoneHeight)
```
and when doing cross-match, one must look *N* zones above and below a zone where
```sql
    N = floor(@theta/@zoneHeight).
```
So, assuming the left and right margins are big enough, the full neighborhood search can be expressed as:
```sql
select objID                                                    -- return the objects
from   ZoneIndex                                                -- from Zone table
where zone between floor((@dec - @theta)/@zoneHeight) --  in the zone
           and floor((@dec + @theta)/@zoneHeight) --  range
  and ra  between @ra  - @alpha and @ra  + @alpha   -- quick filter on ra
  and dec between @dec - @theta and @dec + @theta   -- quick filter on dec
  and (x*@x + y*@y + z*@z) > cos(radians(@theta))   -- careful distance test
```

Unfortunately, the SQL Server optimizer is not smart enough to recognize that it can optimize this plan – it scans all objects in all qualifying zones. So, we give SQL a helping hand. Either by writing a loop and executing the statement for each zone within $\theta$ of the declination dec, or more efficiently for SQL, we create a `Zone` table as:

```sql
create table Zone (
    zone   int not null primary key,  -- floor(latMin/zoneHeight)
    latMin float,                      -- min latitude of this zone (degrees)
    latMax float                       -- max latitude of this zone (degrees)
    )
declare @maxZone bigint, @minZone bigint
set @maxZone =  floor((90.0+@zoneHeight)/@zoneHeight)
set @minZone = - @maxZone
while @minZone < @maxZone              -- Poplate the zone table.
    begin
    insert Zone values (@minZone, @minZone   *@zoneHeight,
                                  (@minZone+1)*@zoneHeight)
    set @minZone = @minZone + 1
    end
```

Then the following query (with its explicit join hint) generates an efficient plan:

```sql
select objID
from (select zone
      from Zone
      where zone between floor((@dec - @theta)/@zoneHeight)
                     and floor((@dec + @theta)/@zoneHeight)) as ZoneHint
inner loop join ZoneIndex on Zone.zone = ZoneIndex.zone
where ra  between ra - @alpha and @ra + @alpha
  and dec between @dec - @theta and @dec + @theta
  and (x*@x + y*@y +z*@z) > cos(radians(@theta))
```

Cross match and self match need a similar table, a `ZoneZone` table that describes all the zones a particular zone must be matched with and also recommends a conservative Alpha to use for all matches in that zone (Alpha is computed knowing the declination and theta.) The Appendix has the definintion of ZoneZone and the code to initialze it.

The crossmatch and self-match then take similar forms (but not identical) The general form is as follows (where the ZoneIndex has been extended with a objType field so that it indexes both datasets.)

```sql
insert CrossMatch
select Z1.objID, Z2.objID,
       degrees(acos(Z1.x*Z2.x + Z1.y*Z2.y + Z1.z*Z2.z)) distance
from            ZoneIndex Z1                            -- from First dataset
inner loop join ZoneZone ZZ  on Z1.zone=ZZ.zone1       -- look in neighbor zones
inner loop join ZoneIndex Z2 on ZZ.Zone2 = Z2.zone    -- at places
where   Z2.ra  between Z1.ra-ZZ.alpha and Z1.ra+ZZ.alpha-- with right longitude
   and Z2.dec between Z1.dec-@theta  and Z1.dec+@theta  -- band
   and Z1.x*Z2.x + Z1.y*Z2.y + Z1.z*Z2.z > cos(radians(@theta)) --
   and Z1.margin = 0                                    -- First not marginal
   and Z1.objType = '1'                                 -- First data set
   and Z2.objType = '2'                                 -- Second data set
```

Again, there are subtle differences between self-match and cross match and some optimization opportunities – but this is the basic idea. The code in the Appendix gives more details showing some additional optimizations – notably exploiting the symmetry of the self-match problem.

## *2.5. Picking an optimal zone height*

As explained in [1], if the typical radius, theta, is known, the optimal zone height is theta. The logic correct derivation was given there.

## 3. Summary

Zones partition an N-Dimensional Euclidian or metric space to efficiently support *points-near-a-point* queries, either within a dataset or between two datasets. The Zones Algorithm uses relational algebra and the B-Tree mechanism found in almost all relational database systems. Zones give a portable-relational implementation of *points-near-point* queries and spatial *cross-match* and *self-match*.

There are a few complications when zones are used in non-Euclidian spaces. In particular, in 2D spherical geometry there is the problem of wrap-around, and the problem that angular distances and coordinates (*lat*, *lon*) or (*ra*,*dec*) must be corrected as the move away from the equator. This article describes fairly simple solutions to both problems. It also points out that the margin logic is subtly different for the three cases of (1) points-near a point, (2) self-match and (3) cross-match. Table 1 summarizes the solutions.

| Table 1. How each issue is treated in each context. | | | |
|---|---|---|---|
| | context | | |
| **Issue** | **Points-near point** | **Self-match** | **Cross-match** |
| **Symmetric test** | Not applicable | Eliminates ½ the work | Not applicable |
| **Spherical Wrap Around** | 180º margins on both sides | *Alpha(theta,dec)* margin on second dataset | |
| **Spherical distance** | *Alpha(theta,dec)* expansion of ra or lat bounding box width, use xyz dot product for careful test | | |
| **Multi-zone** | `Zone` table | `ZoneZone` table | |

The appendix includes a complete implementation of point-near-point, self-match, and cross-match using the USGS city and stream gauge database. The sample code and database can be downloaded from: http://research.microsoft.com/~Gray/zone.zip.

# A. Appendix

## A.1. Defining and Populating The Sample Database

```sql
--------------------------------------------------------------------------------
-- Sample T-SQL code to demonstrate the use of the Zone Algorithm in SQL Server
-- Jim Gray, Alex Szalay, María Nieto-Santisteban
-- Zones.SQL
-- April 2006
--------------------------------------------------------------------------------
-- Create and fills the Zones database from the USGS "Place" and "Station"
-- tables in the SQL 2005 Spatial Samples database (included in the SQL 2005
-- samples.)  You can download this Zone database and attach it rather than
-- run this build script.
--------------------------------------------------------------------------------
set nocount on
create database zones
go
alter database zones set recovery simple
go
use zones
go
--------------------------------------------------------------------------------
-- Place: a USGS list of 22,993 cities in the United States
create table Place (
    PlaceID        int identity not null primary key,
    PlaceName      varchar(100) not null,  -- name of place (e.g. San Francisco)
    State          char(2)      not null,  -- 2 character state code
    Population     int          not null,  -- population circa 1993
    Households     int          not null,  -- households circa 1993
    LandAreaKm     int          not null,  -- area of place
    WaterAreaKm    int          not null,  -- lakes/rivers/ponds in place
    Lat            float        not null,  -- latitude  (degrees)
    Lon            float        not null   -- longitude ((degrees)
    )
create index PlaceName on Place(PlaceName, State)
--------------------------------------------------------------------------------
-- Station: a USGS list of 17,245 stream flow measuring stations in the US
create table  Station(
    StationNumber int           not null primary key, -- USGS ID of station
    StationName   varchar(60)   not null, -- USGS name of station
    State         char(2)       not null, -- 2 character state code
    Lat           float         not null, -- latitude  (degrees)
    Lon           float         not null, -- longitude ((degrees)
    DrainageArea  float         not null, -- area upstream of station
    FirstYear     int           not null, -- when recording started
    YearsRecorded int           not null, -- number of years active
    IsActive      bit           not null, -- is it still active?
    IsRealTime    bit           not null  -- is it online (on the Internet)?
    )
--------------------------------------------------------------------------------
-- populate the Place and Station databases from the SQL 2005
-- Spatial database sample database.
insert Place
     select PlaceName,State,[Population],Households
          ,LandAreaKm, WaterAreaKm, Lat, Lon
     from spatial.dbo.place
insert Station
     select StationNumber,StationName, State, lat, lon, DrainageArea
          ,FirstYear, YearsRecorded, IsActive, IsRealTime
     from spatial.dbo.Station
```

## A.2. Define and Populate the Zone Indices

```sql
--------------------------------------------------------------------------------
-- Create an populate the Zone index tables:
-- ZoneHeight:  contains ZoneHeight constant used by the algorithm.
--              The procedure BuidZoneIndex() populates these tables.
--              You can update ZoneHeight and then call BuidZoneIndex(newHeight)
--              to rebuild the indices
-- ZoneIndex: a table that maps type-zone-longitude to objects
--            it indexes the Place and Station table in this example
-- Zone:      a table of with a row for each zone giving latMin, latMax, Alpha
-- ZoneZone:  Maps each zone to all zones it may have a cross-match with.
--------------------------------------------------------------------------------
-- Use a zone height drives the parameters of all the other tables.
-- Invoke BuidZoneIndex(NewZoneHeight) to change height and rebuild the indices
create table ZoneHeight( [value] float not null) -- zone height in degrees.
--------------------------------------------------------------------------------
-- Zone table has a row for each zone.
-- It is used to force the query optimizer to pick the right plan
create table Zone (
    zone   int not null primary key, -- floor(latMin/zoneHeight)
    latMin float,                    -- min latitude of this zone (degrees)
    latMax float                     -- max latitude of this zone (degrees)
    )
--------------------------------------------------------------------------------
-- Zone-based spatial index for Places and Stations.
-- Note the key is on objectType ('S' or 'P' for station or place in out case)
--               then zone      to give the band to search in
--               then longitude to give an offset in the band.
--               then objID     to give a unique key
-- It copies the spherical and cartesian coordianates from the base objects
-- it also has a flag indicating if this is a "margin" element, to solve
--     the warp-araound problem.
create table  ZoneIndex (
    objType     char(1)     not null, -- P for place, S for station.
    objID       int         not null, -- object Identifier in table
    zone        int         not null, -- zone number (using 10 arcminutes)
    lon         float       not null, -- sperical coordinates
    lat         float       not null,
    x           float       not null, -- cartesian coordinates
    y           float       not null,
    z           float       not null,
    margin      bit         not null, -- "margin" or "native" elements
    primary key (objType, zone, lon, objID)
    )
--------------------------------------------------------------------------------
-- ZoneZone table maps each zone to zones which may have a cross match
create table ZoneZone (
              zone1 int, zone2 int, alpha float,
              primary key (zone1,zone2))
go
--------------------------------------------------------------------------------
-- Function to compute Alpha "expansion" of theta for a given latitude
--   Latitude and theta are in degrees.
create function Alpha(@theta float, @lat float) returns float as
   begin
   if abs(@lat)+@theta > 89.9 return 180
   return(degrees(abs(atan(sin(radians(@theta)) /
                       sqrt(abs(  cos(radians(@lat-@theta))
                                * cos(radians(@lat+@theta))
         )      )   )   )   )   )
    end
```

```sql
--------------------------------------------------------------------------
-- Procedure to populate the zone index.
-- If you want to change the zoneHeight, call this function to rebuild all
-- the index tables.  @zoneHeight is in degrees.
-- @theta is the radius of cross-match, often @theta == @zoneHeight
create procedure BuidZoneIndex(@zoneHeight float, @theta float) as
    begin
    --------------------------------------------------------------------
    -- first empty all the existing index tables.
    truncate table ZoneHeight
    truncate table Zone
    truncate table ZoneIndex
    truncate table ZoneZone
    --------------------------------------------------------------------
    -- record the ZoneHeight in the ZoneHeight table
    insert ZoneHeight values (@zoneHeight)
    --------------------------------------------------------------------
    -- fill the zone table (used to help SQL optimizer pick the right plan)
    declare @maxZone bigint, @minZone bigint
    set @maxZone =  floor((90.0+zoneHeight) /@zoneHeight)
       set @minZone = - @maxZone
    while @minZone < @maxZone
        begin
        insert Zone values (@minZone, @minZone   *@zoneHeight,
                                      (@minZone+1)*@zoneHeight)
        set @minZone = @minZone + 1
        end
    --------------------------------------------------------------------
    -- Create the index for the Place table.
    Insert ZoneIndex
      select 'P', PlaceID,
             floor((lat)/@zoneHeight) as zone,
             lon, lat,
             cos(radians(lat))*cos(radians(lon)) as x,
             cos(radians(lat))*sin(radians(lon)) as y,
             sin(radians(lat)) as z,
             0 as margin
      from Place
    --------------------------------------------------------------------
    -- Create the index for the Station table.
    Insert ZoneIndex
      select 'S', StationNumber,
             floor((lat)/@zoneHeight) as zone,
             lon, lat,
             cos(radians(lat))*cos(radians(lon)) as x,
             cos(radians(lat))*sin(radians(lon)) as y,
             sin(radians(lat)) as z,
             0 as margin
      from Station
    --------------------------------------------------------------------
    -- now add left and right margin
    -- You could limit the margin width use Alpha(MaxTheta,zone.maxlat)if you
    --  knew MaxTheta;  but, we do not know MaxTheta so we use 180
    Insert ZoneIndex
      select [objType], objID, zone,
           lon-360.0, lat, x, y, z,
           1 as margin             -- this is a marginal object
      from ZoneIndex where lon >= 180 -- left margin
    union
      select [objType], objID, zone,
           lon+360.0, lat, x, y, z,
           1 as margin             -- this is a marginal object
      from ZoneIndex where lon < 180  -- right margin
```

```sql
      ---------------------------------------------------------------------
      -- ZoneZone table maps each zone to zones which may have a cross match
      declare @zones int       -- number of neighboring zones for cross match
      set @zones = ceiling(@theta/@zoneHeight) -- (generally = 1)
      insert ZoneZone          -- for each pair, compute min/max lat and Alpha
         select Z1.zone, Z2.zone, case when Z1.latMin < 0
                                       then dbo.Alpha(@theta, Z1.latMin)
                                       else dbo.Alpha(@theta, Z1.latMax) end
         from Zone Z1 join Zone Z2
         on Z2.zone between Z1.zone - @zones and Z1.zone + @zones
end
go
------------------------------------------------------------------------------
-- Initial call to build the zone index with a height of 10 arcMinutes.
-- and a self-match or cross-match radius of 1 degree (60 nautical miles).
declare @zoneHeight float, @theta float
set @theta      = 60.0 / 60.0
set @zoneHeight = 10.0 / 60.0
exec BuidZoneIndex  @zoneHeight, @theta
go
```

## *A.3. Define And Use Points-Near-Point Function*

```sql
--------------------------------------------------------------------------------
-- GetNearbyObjects() returns objects of type @type in { 'P', 'S'}
-- that are within @theta degrees of (@lat, @lon)
-- The returned table includes the distance to the object.
create function GetNearbyObjects(
                    @type char(1),          -- 'P' or 'S'
                    @lat float, @lon float, --  in degrees
                    @theta float)           -- radius in degrees
returns @objects Table (objID int primary key, distance float) as
    begin
    declare @zoneHeight float, @alpha float,
            @x float, @y float, @z float
    -- get zone height from constant table.
    select  @zoneHeight = min([value]) from ZoneHeight
    -- compute "alpha" expansion and cartesian coordinates.
    select  @alpha = dbo.Alpha(@theta, @lat),
            @x = cos(radians(@lat))*cos(radians(@lon)),
            @y = cos(radians(@lat))*sin(radians(@lon)),
            @z = sin(radians(@lat))
    -- insert the objects in the answer table.
    insert @objects
    select objID,
           case when(@x*x +@y*y + @z*z) < 1   -- avoid domain error on acos
                then degrees(acos(@x*x +@y*y + @z*z))
                else 0 end                    -- when angle is tiny.
       from Zone Z                             -- zone nested loop with
       inner loop join ZoneIndex ZI on Z.zone = ZI.zone -- zoneIndex
       where objType = @type                   -- restrict to type 'P' or 'S'
         and Z.latMin between @lat-@theta-@zoneHeight  -- zone intersects
                          and @lat+@theta     -- the theta circle
         and ZI.lon between @lon-@alpha       -- restrict to a 2 Alpha wide
                    and    @lon + @alpha      -- longitude band in the zone
         and ZI.lat between @lat-@theta       -- and roughly correct latitude
                    and    @lat + @theta
         and (@x*x +@y*y + @z*z)              -- and then a careful distance
                    > cos(radians(@theta))    -- distance test
    return
    end
go
--------------------------------------------------------------------------------
-- GetNearstObject() returns the object of type @type in { 'P', 'S'}
-- nearest to (@lat, @lon)
create function GetNearestObject(
                    @type char(1),          -- 'P' or 'S'
                    @lat float, @lon float) --  in degrees
returns @objects Table (objID int primary key, distance float) as
    begin
    declare @theta float                  -- uses GetNearbyObjects
    set @theta = .2                       -- with radius starting at 12 nautical
    while(1=1)                            -- miles and increasing 2x on each
        begin                             -- probe till a  hit is found
        insert @objects                   -- put top 1 (== shortest distance)
        select top 1 objID, distance   -- object in the target table.
           from GetNearbyObjects('P', @lat, @lon, @theta)
           order by distance desc
        if @@rowcount != 0 break          -- stop when  select finds something
        set @theta = @theta*2             -- otherwise double the search radius
        end                               --
   return                                 -- return closest object and its distance
end
```

```sql
go
---------------------------------------------------------------
-- Three test cases:
declare @lat float , @lon float, @theta float
set @theta =  .2      -- 30 nautical miles == 1/2 degree
set @lat =   37.8   -- the approximate center of San Francisco
set @lon = -122.56
-- find cities nearby San Francisco.
select str(60*distance,5,1) as distance,
       cast(PlaceName+', '+ state as varchar(30)) place,
       population, households, landAreaKm, WaterAreaKm,
       str(Lat,8,4) Lat, str(Lon,10,4) Lon
from GetNearbyObjects('P',@lat, @lon, @theta) O join Place P on ObjID = PlaceID
order by distance
/* returns:
distance place                          population households landAreaKm WaterAreaKm Lat      Lon
-------- ------------------------------ ---------- ---------- ---------- ----------- -------- ----------
  0.5    San Francisco, CA              723959     328471     309644     1227875     37.7933  -122.5548
  4.7    Sausalito, CA                  7152       4378       12400      2567        37.8577  -122.4915
  5.4    Tamalpais-Homestead Valley, CA 9601       4251       19159      81          37.8884  -122.5390
  6.1    Belvedere, CA                  2147       1037       3548       12508       37.8717  -122.4687
  6.3    Strawberry, CA                 4377       2241       8842       99          37.8970  -122.5077
  6.5    Mill Valley, CA                13038      6139       30970      808         37.9080  -122.5410
  7.1    Tiburon, CA                    7532       3433       28948      57914       37.8868  -122.4568
  7.5    Broadmoor, CA                  3739       1274       2941       0           37.6917  -122.4796
  7.8    Corte Madera, CA               8272       3717       21027      8243        37.9236  -122.5073
  8.1    Daly City, CA                  92311      30162      49820      0           37.6870  -122.4674
  8.6    Larkspur, CA                   11070      5966       20787      896         37.9412  -122.5292
  9.0    Kentfield, CA                  6030       2492       15902      33          37.9503  -122.5474
  9.1    Colma, CA                      1103       437        12608      0           37.6738  -122.4534
  9.1    Bolinas, CA                    1098       584        9108       0           37.9056  -122.6970
  9.7    Ross, CA                       2123       768        10572      0           37.9618  -122.5606
 10.1    Brisbane, CA                   2952       1382       22095      898         37.6893  -122.3999
 11.0    San Anselmo, CA                11743      5330       18432      0           37.9826  -122.5689
 11.2    San Rafael, CA                 48404      21139      109967     38727       37.9811  -122.5059
 11.4    Fairfax, CA                    6931       3225       13885      0           37.9886  -122.5938
*/
-- find stream gauge closest to San Francisco.
select str(60*distance,5,1) as distance,
       cast(StationName as varchar(40)) Station
from GetNearestObject('S',@lat, @lon) O
join Station S on  ObjID = StationNumber
/* returns:
distance Station
-------- ----------------------------------------
 11.3    Spruce Branch A South San Francisco Ca
*/
-- find stream gauges near to San Francisco.
select str(60*distance,5,1) as distance,
       cast(StationName as varchar(35)) StationName,
       FirstYear, YearsRecorded,
       str(Lat,8,4) Lat, str(Lon,10,4) Lon
from GetNearbyObjects('S',@lat, @lon, @theta) O
join Station S on ObjID = StationNumber
/* returns:
distance StationName                            FirstYear  YearsRecorded Lat      Lon
-------- -------------------------------------- ---------- ------------- -------- ----------
 10.8    Colma C A South San Francisco Ca       1964       31            37.6538  -122.4274
 11.3    Spruce Branch A South San Francisco    1965       5             37.6460  -122.4230
 10.4    San Rafael C A San Rafael Ca           1972       2             37.9726  -122.5375
  9.7    Corte Madera C A Ross Ca               1951       43            37.9623  -122.5577
  5.9    Arroyo Corte Madera D Pres A Mill V    1966       20            37.8971  -122.5372
  8.9    Morses C A Bolinas Ca                  1967       3             37.9190  -122.6713
  9.5    Pine C A Bolinas Ca                    1967       4             37.9185  -122.6941
*/
```

## A.4. Cross Match Places with Stations and Places with Places

```
------------------------------------------------------------------------
-- CROSS MATCH EXAMPLE and SELF-MATCH EXAMPLE
------------------------------------------------------------------------
declare @theta float, @zoneHeight float
set     @theta = 60.0 / 60.0
-- optionally change the zone height
-- exec BuidZoneIndex  @theta  -- @theta is the best zone height for cross
                               -- match. but we do not need to change zone
                               -- height to make the algorthim below work
------------------------------------------------------------------------
-- CROSS MATCH EXAMPLE
------------------------------------------------------------------------
-- cross match places with stations using a 60 nautical mile radius
create table PlaceStationCrossMatch(
    PlaceID        int    not null,       -- ID of place (e.g. San Francisco)
    StationNumber  int    not null,       -- ID of station (e.g. Bolinas)
    distanceNM     float  not null)       -- distance to station from place
--  primary key (PlaceID, StationNumber)) -- primary key added later
------------------------------------------------------------------------
-- Compute the cross match between Places and stations.
insert PlaceStationCrossMatch
select P.objID placeID, S.objID stationID,
       60* degrees(acos(P.x*S.x + P.y*S.y + P.z*S.z)) distanceNM
from           ZoneIndex P                       -- start with a place
inner loop join ZoneZone ZZ on P.zone=ZZ.zone1   -- look in neighbor zones
inner loop join ZoneIndex S on ZZ.Zone2 = S.zone -- at places
where   S.lon between P.lon-ZZ.alpha and P.lon+ZZ.alpha -- with right longitude
    and S.lat between P.lat-@theta  and P.lat+@theta  -- band
    and P.x*S.x + P.y*S.y + P.z*S.z > cos(radians(@theta)) -- distance test
    and P.margin = 0                              -- place not marginal
    and P.objType = 'P'                           -- First object is a place
    and S.objType = 'S'                           -- Second object is a station
-- 29 seconds,  2,476,665 objects (19 seconds for just the computation)
------------------------------------------------------------------------
-- now add primary key
alter table PlaceStationCrossMatch
add constraint pk_PlaceStationCrossMatch
primary key clustered (PlaceID, stationNumber)
-- 18 seconds.
```

```sql
--------------------------------------------------------------------------
-- SELF-MATCH EXAMPLE
--------------------------------------------------------------------------
-- cross match Places with Places using a 60 nautical mile radius
-- This uses the symmetric cross-match (doing 1/2 the work in in step 1)
create table PlacePlaceCrossMatch(
        PlaceID         int   not null,  -- ID of place (e.g. San Francisco)
        PlaceID2        int   not null,  -- ID of nearby place (e.g. Bolinas)
        distanceNM      float not null ) -- distance to station from place
--  primary key (PlaceID, PlaceID2)) -- primary key added after table built
--------------------------------------------------------------------------
-- Do the zone-zone cross match -- since it is self-match can do first 1/2
-- by objID1 < objID2.
insert PlacePlaceCrossMatch
select P1.objID placeID, P2.objID PlaceID2,
        60* degrees(acos(P1.x*P2.x + P1.y*P2.y + P1.z*P2.z)) distanceNM
from           ZoneIndex P1                             -- for each place
inner loop join ZoneZone Z   on P1.zone  = Z.zone1      -- look in nearby zones
inner loop join ZoneIndex P2 on  Z.Zone2 = P2.zone      -- look at other places
where   P2.lon between P1.lon-Z.alpha and P1.lon+Z.alpha-- in right longitude
        and P2.lat between P1.lat-@theta  and P1.lat+@theta -- in the right lat
        and P1.x*P2.x + P1.y*P2.y + P1.z*P2.z           --right distance
                                > cos(radians(@theta)) --
    And P1.margin = 0                                   -- first not marginal
    and P1.objID < P2.objID                             -- the 50% test
    and P1.objType = 'P'                                -- both are places
    and P2.objType = 'P'
-- 37 seconds,  2,594,621 objects (19 seconds for just the computation)
--------------------------------------------------------------------------
-- Do the mirror image (the other 2.6 M rows) in 33 seconds
insert PlacePlaceCrossMatch
select PlaceID2, placeID, distanceNM                    -- note placeID reversal
from PlacePlaceCrossMatch                               --
-- 40 seconds
--------------------------------------------------------------------------
-- build the clustering index on the resulting table. 5.2 M rows in 18 seconds
alter table PlacePlaceCrossMatch
add constraint PK_PlacePlaceCrossMatch
primary key clustered (PlaceID, PlaceID2)
--------------------------------------------------------------------------
-- 27 seconds
```